\documentclass[aps, onecolumn,  showpacs, amsmath,  amssymb, prd, floatfix, 10pt]{revtex4} 
\usepackage{graphicx}     
\usepackage{natbib}

\begin{document}
  
\title{ Kobayashi-Maskawa  matrix moduli, decay constants and form factors determination  from  experimental data}
\author{Petre Di\c t\u a$^*$}
\affiliation{ National Institute of Physics and Nuclear Engineering\\
P.O. Box MG6, Bucharest, Romania}
\date{October 14, 2010}

\begin{abstract}
The aim of the paper is to propose another  tool for phenomenological analyses  of experimental data from superallowed  nuclear and neutron  $\beta$ decays, and  from leptonic and semileptonic decays, that allows the  finding of the most probable numerical form of the CKM matrix, as well as the determination of decay constants, $f_{P}$, and of various form factors  $f_+(q^2)$, by using another implementation of unitarity constraints. In particular this approach allows  the determination of semileptonic form factors that is illustrated on the existing data from $D\rightarrow \pi l \nu$ and $D\rightarrow K l\nu$ decays.
\end{abstract}


\maketitle

\section{Introduction}
Within the Standard Model (SM) the flavor physics is encoded by  Kobayashi-Maskawa (KM) matrix, \cite{KM}, supposed to be unitary, matrix that  describes the quark flavor mixing through  four independent parameters: three mixing angles, $\theta_{ij}$, $ij$ = 12, 13, 23, and one {\em CP}-violating phase, $\delta$. By consequence the experimental determination of KM matrix entries is essential for the validation of the SM, and for  detection of new physics beyond it. 

However the determination of KM entries is not an easy problem because of two different causes. The first one is theoretical, namely  the mixing angles are not invariant quantities, their numerical values depend on the original KM form,   \cite{KM}, or on the present day form,  \cite{pdg08}, which  is not rephasing invariant, see  \cite{HL}. 
 These shortcomings  are harmless if one follows Jarlskog's solution. Starting with her first papers on  KM matrix,  \cite{J1},  she  proposed the determination of 
 the quark mixing matrix in terms of directly measurable quantities, and, in the same time, invariant quantities. In this context an invariant quantity is one whose numerical value does not depend on the KM matrix form, or of its rephaising invariance.   Jarlskog  provided two such invariants:  the KM moduli, and the celebrated  $J$ invariant, \cite{J2}. Ten years later, other invariants,  the angles of unitarity triangles, 
appeared on  the scene \cite{AKL}. After few years it was realized that all the measurable quantities of  quark mixing matrix are expressible in terms of four independent KM matrix moduli, see \cite{BL} and \cite{J3}. In these papers it was also shown that  areas of all unitarity triangles are equivalent, and  numerically equal to half of $J$ invariant, and in our opinion this equality also solved the   $J$ sign  problem, by imposing  $J > 0$.
As a conclusion we can state that the numerical values for all measured invariant quantities  should be the same irrespective of the physical processes  where they are involved.

The second difficulty comes from the experimental side. If one wants the use of KM moduli as independent parameters these ones are not directly measured by experimenters. In the simplest case, that of leptonic decays, they measure branching ratios and provide numbers for  products of the form $|U_{qq'}|f_P$, where $U_{qq'}$ is the corresponding KM matrix element, and $f_P$ is the decay constant. For meson semileptonic decays the physical observable is the differential decay rate, $d\Gamma/dq^2$, which up to known factors, is proportional to  $|U_{qq'}f_+(q^2)|^2$, where $f_+(q^2) $ is a complex form factor and $q$ denotes the transferred momentum between initial and final  mesons. In the last case the experimental teams usually provide numerical values for  products of the form $|U_{qq'}f_+(0)|$, where $f_+(0)$ is the semileptonic decay form factor at zero-momentum transfer. It is clear that from such  measurements one cannot find two unknowns, say  $|U_{qq'}|$ and $f_P$. This can be done if  and only if one can find independent constraints on KM matrix moduli; fortunately  these ones are provided by unitarity.
Thus the main aim of the paper is to show how  the  unitarity property of the KM matrix can be transformed into a powerful tool for the determination  of both matrix moduli and form factors directly from experimental data.

The unitarity constraints are presented in Sec. II  where they are implemented in a $\chi^2-$form that depends only on KM matrix moduli. In Sec. III we present the decay formulas for superallowed $0^+\rightarrow 0^+$ nuclear and neutron  $\beta$ decays, and those for leptonic and semileptonic decays. They depend on KM matrix  moduli and  specific decay parameters such as decay constants, $f_P$, and form factors $|f(q^2)|$, etc, that are implemented in an other $\chi^2$-piece. In Sec. IV we cite the experimental papers from which  we take data  that are used in our fit, and in Sec. V we present numerical results. The paper ends by Conclusion.

\section{Unitarity constraints}

The use of $|U_{ij}|$ as independent parameters raises an important problem.
This means that we have to solve  the consistency problem between moduli and unitarity property, which all amounts to obtaining the necessary and sufficient conditions on  the set of numbers $|U_{ij}|$  to represent the moduli of an exact unitary matrix. After that we have to find a device for applying these conditions   to  experimental situation where  data are known modulo uncertainties. 

Both these problems  were recently solved, and a procedure for recovering   KM matrix elements from error affected data was provided in \cite{PD}. These unitarity constraints say that the four independent parameters  $s_{ij}=\sin\theta_{ij}$ and $\cos\delta$  should take physical values, i.e. $s_{ij}\in (0,1)$ and $\cos\delta \in (-1,1)$, when they are computed via   equations set: 
\begin{eqnarray}
V_{ud}^2&=&c^2_{12} c^2_{13},\,\, V_{us}^2=s^2_{12}c^2_{13},\,\,V_{ub}^2=s^2_{13}\nonumber \\
 V_{cb}^2&=&s^2_{23} c^2_{13},\,\,
 V_{tb}^2=c^2_{13} c^2_{23},\nonumber\\
V_{cd}^2&=&s^2_{12} c^2_{23}+s^2_{13} s^2_{23} c^2_{12}+2 s_{12}s_{13}s_{23}c_{12}c_{23}\cos\delta,\nonumber\\
V_{cs}^2&=&c^2_{12} c^2_{23}+s^2_{12} s^2_{13} s^2_{23}-2 s_{12}s_{13}s_{23}c_{12}c_{23}\cos\delta,~~~~~\label{uni}\\
V_{td}^2&=&s^2_{13}c^2_{12}c^2_{23}+s^2_{12}s^2_{23}-2 s_{12}s_{13}s_{23}c_{12}c_{23}\cos\delta\nonumber,\\
V_{ts}^2&=&s^2_{12} s^2_{13} c^2_{23}+c^2_{12}s^2_{23} +2 s_{12}s_{13}s_{23}c_{12}c_{23}\cos\delta\nonumber
\end{eqnarray}
The above relations have been  obtained by using the standard KM matrix form, \cite{pdg08}, where $V_{ij}=|U_{ij}|$, and $U_{ij}$ are    KM matrix  entries. In paper \cite{PD} it was  shown  that if the independent parameters are KM matrix moduli the reconstruction of a unitary matrix  knowing its moduli  is essentially unique. By consequence in the following the used independent parameters in all our phenomenological analyses  will be  $V_{ij}$ moduli. Although only four of them can be independent the experimental data ``force'' us to use all the possible sets of four independent moduli, as it will be shown in the following, and a simple combinatorial evaluation shows that there are 57 such sets.

Relations (\ref{uni}) are rephaising invariant, i.e. they have the same form after    multiplication of  all  KM matrix rows and columns   by arbitrary phases. More important is that  they   contain all unitarity constraints.  It is easily seen that all the six relations such as 
\begin{eqnarray}
V_{ud}^2+V_{us}^2+V_{ub}^2=1\label{ds}
\end{eqnarray}
are a consequence of the above relations. The  relations (\ref{ds}) are necessary conditions for  unitarity fulfilment, but not \textit{sufficient} as we show in the following. In fact relations  (\ref{ds}) show that if they are satisfied then there exist a physical solution for $s_{ij}$.
Indeed
from any four independent moduli entering  (\ref{uni}) one can get the mixing parameters  $s_{ij}$ and $\cos\delta$, i.e. the four independent  parameters entering KM unitary matrix.
 For example, if $V_{us}=a,\,V_{ub}=b,\,{\rm and}\,\,V_{cb}=c$ is one  set of three independent moduli, from the first five equations  (\ref{uni}) we find all the three mixing  parameters
\begin{eqnarray}
s_{13}=V_{ub}=b,\,\,s_{12}=\frac{a}{\sqrt{1-b^2}}, \,\,s_{23}=\frac{c}
{\sqrt{1-b^2}}\label{sol}\end{eqnarray}
The other parameter, $\delta$, can be obtained   from anyone of the last four equations. If we choose the sixth equation one gets
\begin{eqnarray}
\cos\delta = 
\frac{(1-b^2)(V_{cd}^2(1-b^2)-a^2)+c^2(a^2+b^2(a^2+b^2-1))}{2 a b c \sqrt{1-a^2-b^2}\sqrt{1-b^2-c^2}\label{col}}
\end{eqnarray}
and from the remaining relations three new  $\cos\delta$ formulas similar to (\ref{col}). 
The above relation shows that $\cos\delta$ is an other invariant in the Jarlskog sense depending of four independent moduli, and {\em CP}-violation phase can be measured via relations such as (\ref{col}).

If we make use of the last four relations (\ref{uni}) we get only one solution for mixing parameters and $\cos\delta$. Thus  depending on the chosen four independent moduli set the number of solutions varies between one and four. 
Because there are 57 such  groups one get 165 different expressions for $\cos\delta$. They  take  the same numerical value when are  computed via Eqs. (\ref{uni}), {\em if and only if all the six relations similar to Eq.(\ref{ds}) are exactly satisfied}. If the moduli matrix generated by four independent moduli is compatible with unitarity then  $\cos\delta\in(-1,1)$, and outside this  interval when the corresponding matrix is not compatible. For example if we choose
$V_{us}=2257/10^4,\;V_{ub}=359/10^5,\;V_{cd}=2256/10^4,\; {\rm and}\; V_{cb}=415/10^4$, by using the necessary relations similar to (\ref{ds}),  the corresponding moduli matrix is
\begin{eqnarray}|U|=\left(\begin{array}{ccc}
\frac{\sqrt{9490466219}}{10^5}&\frac{2257}{10^4}&\frac{359}{10^5}\\*[2mm]
\frac{141}{625}&\frac{3\sqrt{10526471}}{10^4}&\frac{83}{2\times 10^3}\\*[2mm]
\frac{\sqrt{580181}}{10^5}&\frac{\sqrt{10982}}{2500}&\frac{\sqrt{9982648619}}{10^5}
\end{array}\right)\label{pdg}\end{eqnarray}
and from (\ref{col}) one gets $\cos\delta \approx 0.64088$, showing that the above moduli matrix, (\ref{pdg}), comes from  an exact unitary matrix.

If we modify the previous numerical $V_{us}$ value  by adding to it the small quantity $3\times 10^{-4}$ the mixing parameters are still physical, only $s_{12}$ is modified by a very small quantity, and respectively all the square root entries of (\ref{pdg}), necessary for fulfilment of all the six relations similar to (\ref{ds}).  In this case one gets
$\cos\delta\approx-1.42427$, which shows that the new  moduli matrix, $|U|$,  is not compatible with unitarity, even  it \textit{exactly} satisfies all the six relations (\ref{ds}). 

  If one computes the $J$ invariant one finds in the  above two cases
 \begin{eqnarray}J^2=6.317 \times10^{-10},\;{\rm and}\;  J^2=-1.106 \times10^{-9}\end{eqnarray}
Thus the physical conditions for unitarity compatibility are $\cos\delta\in (-1, 1)$, and $J^2 >  0$,  respectively, and from a theoretical point of view they are equivalent. For numerical computations the use of $\cos\delta$ formulas, like (\ref{col}), seems to be more efficient because of their great  sensitivity to small moduli variation.

The real physical cases are those where the central value moduli matrices, directly determined from data, or from a fit do not  exactly satisfy   relations (\ref{ds}), but only approximately; for example for a good fit the difference could be  $10^{-5}-10^{-7}$, i.e. rather small from a phenomenological point of view. In these cases the different formulas for  $\cos\delta$ provide different values, physical and unphysical,  even if the mixing parameters take physical values as in previous example. Hence physical reality obliges us to implement the unitarity  constraints
 \begin{eqnarray}
 \cos\delta_i\approx\cos\delta_j,\;i\ne j,\; {\rm all}\; \cos\delta_i\in(-1,1) 
\end{eqnarray}
into a $\chi^2$ fitting device, and our choice is 
\begin{eqnarray}
\chi^2_{1}=
\sum_{j=u,c,t}\left(
\sum_{i=d,s,b}V_{ji}^2-1\right)^2
+\sum_{j=d,s,b}\left(
\sum_{i=u,c,t}V_{ij}^2-1\right)^2
+ \sum_{i < j}(\cos\delta^{(i)} -\cos\delta^{(j)})^2,\,\,\,\,-1\le\cos\delta^{(i)}\le 1 \label{chi1}
\end{eqnarray}
 that enforces all  unitarity constraints.

\section{Decay Formalism}

In this section we present the decay formalism such as it is used for the description of available experimental data, formalism that  allows us to define a second piece of the $\chi^2$-function by taking into account as much as possible the physical information.

 Information on $V_{ud}$ come from two important sources, superallowed $0^+\rightarrow 0^+$ nuclear $\beta$ decay, and neutron $\beta$ decay. Superallowed $0^+\rightarrow 0^+$  $\beta$ decay between $T = 1$ analog states depends uniquely on the vector part of the weak interaction and, according to the conserved vector current  hypothesis, its experimental $f t$ value should be related to the vector coupling constant, which is a fundamental constant and by consequence has the same value for all such transitions, see \cite{HT1}, \cite{HT2}, \cite{HT3} and \cite{GS}. This means that the following relation should hold
\begin{eqnarray}
ft=\frac{K}{2|G_V|^2\,|M_F|^2}= \textrm{const}\label{g1}
\end{eqnarray}
where $K/(\hbar c)^6=2\pi^3\,\hbar\,{\rm ln 2}/(m_e c^2)^5$, $G_V$ is
the vector coupling constant for semi-leptonic weak interactions, and
$M_F$ is the Fermi matrix element which in this  case is equal to $\sqrt{2}$.
The 
$ft$ value that characterizes any $\beta$ transition depends on the total transition energy $Q_{EC}$,  the  half-life, $t_{1/2}$,  of the parent state, and the branching ratio for the particular studied transition, \cite{HT2}. The above relation is only approximately satisfied by a restricted data set, and for this set  one defines a ``corrected'' value $\mathcal{F}t\equiv ft(1+\delta^{'}_{R})(1+\delta_{NS}-\delta_C) $, where $\delta^{'}_{R} $ and  $\delta_{NS} $ comprise the transition-dependent part of the radiative correction, while  $\delta_C$ depends on the details of nuclear structure. In the above formula we take $|G_V^2|=g_V^2 V_{ud}^2$, with  $g_V =1$, and  write it  as
\begin{eqnarray}
\mathcal{F} t =\frac{K}{2  V_{ud}^2(1+\Delta_R^V)}\label{g2}\end{eqnarray}
where  $\Delta_R^V $ is the  transition-independent part of the radiative corrections whose last estimation given in \cite{HT2} is 
\begin{eqnarray}\Delta_R^V =(2.361\pm0.038)\% \label{rad}\end{eqnarray}

Similarly for neutron $\beta$ decay we make use of  formula
\begin{eqnarray}
V_{ud}^2(1+3\lambda^2)=\frac{4908.7(1.9)s}{\tau_n}\end{eqnarray}
see \cite{I}, where $\tau_n$ is the neutron mean life and $\lambda=g_A/g_V$. In our approach $V_{ud},\;\Delta_R^V,\;{\rm and}\; \lambda$  are free parameters to be found from fit.

In  SM the purely leptonic decay of a $P$ meson, $P\rightarrow l \bar{\nu_l}$, proceeds via annihilation of  the quark pair  to a charged lepton and neutrino through exchange of a virtual $W$ boson, and the branching fraction, up to radiative corrections,   has the form 
\begin{eqnarray}
{\mathcal{B}}(P\rightarrow  l \bar{\nu_l})=
\frac{G_F^2M_Pm_l^2}{8\pi \hbar}\left(1-\frac{m_l^2}{M_P^2}\right)^2f_P^2 V_{qq'}^2 \tau_P\end{eqnarray}
where $G_F$ is the Fermi constant, $M_P$ and $m_l$  are  the $P$ meson  and  $l$ lepton masses, respectively, $f_P$ is the decay constant, $V_{q q'}$ is the modulus of the corresponding KM matrix element, and $\tau_P$ is  $P$ lifetime.

 The next simple decays involving $V_{ij}$ moduli are 
 the semileptonic decays of   heavy pseudoscalar mesons, $H$, into  lighter ones, $P$, whose physical observable   
 is the differential decay rate,  written as
\begin{eqnarray}
\frac{d\,\Gamma(H\rightarrow P\,\ell\,\nu_{\ell})}{d q^2}=\frac{G_F^2\,V_{qq'}^2}{192\pi^3 M_H^3}\lambda^{3/2}(q^2)|f_+(q^2)|^2\label{g4}
\end{eqnarray}
where $q=p_H-p_P$ is the transferred momentum,
and $f_+(q^2)$ is  the global form factor which is a combination of  the two form factors generated by the vector part of the weak current.
When the leptons are electrons, or muons whose masses are low compared to mass difference $m_H-m_P$, $\lambda(q^2)$ is the usual triangle function
\begin{eqnarray}\lambda(q^2)=(M_H^2+M_P^2-q^2)^2-4 M_H^2 M_P^2 \end{eqnarray}

For the decay $\bar{B}\rightarrow Dl\nu$ the experimenters make use of an other variable, namely ${\it w}=(M_B^2+M_D^2-q^2)/(2M_BM_D)$. When the $\tau$ lepton is involved  the above formulas are a little bit more complicated, see \cite{CLN}.

 Usually, till now, the experimenters provided numerical values for  products of the form $V_{qq'}\,f_+(0)$, but we expect that   data of the form $V_{qq'}\,|f_+(q^2)| $ will be available in the near future for all semileptonic decays. 

The second $\chi^2-$component 
which takes into account the experimental data has the form 
\begin{eqnarray}
\chi^2_2=\sum_{i}\left(\frac{d_{i}-\widetilde{d}_{i}}{\sigma_{i}}\right)^2\label{chi2} 
\end{eqnarray}
where  $d_{i}$ are the theoretical functions one wants to be found from fit, $\widetilde{d}_{i}$ are  the numerical values that describe the corresponding experimental data, while  $\sigma_i$ is the  uncertainty  associated to $\widetilde{d}_{i}$. For semileptonic decays $d_i$ could be of the form  $d_i= |f_+(q^2_i)|V_{kl}$ .
 In the following our fitting  $\chi^2$-function will be 
\begin{eqnarray}
\chi^2=\chi^2_1 +\chi^2_2\label{chi}\end{eqnarray}

As it is easily seen  $V_{ij}$ moduli enter naturally in all  formulas that describe leptonic and semileptonic decays being, in our opinion, a strong argument 
for their use as fit parameters.

\section{Experimental Data}

In our analysis we used superallowed $0^+  \rightarrow 0^+$ nuclear $\beta$ decays from \cite{HT1}-\cite{GS}, and the neutron lifetime  from four papers: \cite{AS}, \cite{JN}, \cite{MD}, \cite{SAr}, for $V_{ud}$ determination. We also used  four values for the  $\beta$-asymmetry parameter  $A_0$, from papers \cite{RP}, \cite{HA}, 
\cite{HA1}, \cite{BY}, and one  for the electron-antineutrino correlation coefficient $a_0$, \cite{JB}, for $\lambda$ determination.

$V_{us}$ modulus is involved in kaon and pion leptonic and semileptonic decays, but also in ratio $V_{us}/V_{ud}$, as  for example in Marciano relation, \cite{WJM}, that we write as
\begin{eqnarray}
\frac{V_{us}^2f_K^2}{V_{ud}^2f_{\pi}^2}(1+C_r)=
\frac{{\mathcal{B}}(K\rightarrow \mu\bar{\nu}_{\mu}(\gamma))\tau_{\pi}m_{\pi}(1-\frac{m_{\mu}^2}{m_{\pi}^2})^2}{{\mathcal{B}}(\pi\rightarrow \mu\bar{\nu}_{\mu}(\gamma))\tau_{K}m_K(1-\frac{m_{\mu}^2}{m_{K}^2})^2}
\end{eqnarray}
where $C_r$ is a radiative correction stemming from both $\pi$ and $K$ hadronic structures.

 The last  numerical values for the product $f_+^{\pi K}(0)V_{us}$ are given by KLOE collaboration,   \cite{Kl}, and by FlaviaNet Working Group, \cite{Fl}, and in fit we made use of all the (little) different $f_+(0)V_{us}$ values corresponding to  the five channels. We also used  results from \cite{MT}, \cite{JR}, \cite{CB}, \cite{FA}, \cite{Ts}, \cite{Vr}, \cite{Mm}, \cite{FA2}, \cite{FA3}, \cite{MT1}, \cite{AB}, \cite{TA} that give only a ``mean value'' for the above product.

 $V_{ub}$ is  the most poorly determined modulus although there is much experimental information coming from  decays $B\rightarrow \pi l\nu$, see \cite{SBA}, \cite{TH}, \cite{KI}, \cite{NEA},  \cite{BA2}, \cite{FN}, \cite{BA3}. In this papers the experimenters have been confronted with the known difficulty,  getting two distinct parameters, $V_{ub}$ and $ |f_+^{B\pi}(q^2)|$, from their product measured from experimental data. Thus they used the form factor lattice computations to obtain  $V_{ub}$ values depending on $q^2$, see \cite{SBA}, \cite{TH}, and \cite{BA3}. For fit  we found three  measurements for  $f_BV_{ub}$, 
\cite{KI}, \cite{BA1}, \cite{BA2}, and three fenomenological determinations involving
 $f_+^{B\pi}(0)V_{ub}$, \cite{FN}, \cite{BA3}, \cite{BLC}.

$V_{cd}$ and $V_{cs}$ moduli enter the leptonic decays  $D\rightarrow l\nu$, \cite{BE}, \cite{AR}, \cite{ID}, and $D_s^+\rightarrow l\nu$, \cite{JPA}, \cite{PU}, \cite{KME}, \cite{LW}, \cite{AR1}, respectively, as well  as the semileptonic decays  $D\rightarrow \pi l\nu$, and $D\rightarrow K l\nu$, \cite{DCH}, \cite{SD}, \cite{JG}. In the last three papers one find $V_{cq} f_+(0)$
 values, for $q=d$, or  $s$,  and for the first time values on 
\begin{eqnarray}
|f_+(q^2)|V_{cq}=\sqrt{\frac{d\Gamma}{dq^2}\frac{24 \pi^3 p^3_{K,\pi}}{G_F^2}}\label{ff}
\end{eqnarray}
\cite{JG}, giving the possibility of extracting    form factors directly from data. The above semileptonic decays  allow the measurement of the ratio
$V_{cd}f_+^{D\pi}(0)/V_{cs}f_+^{DK}(0)$, see \cite{SD}, \cite{JG}, \cite{LW1}, \cite{GSH}, \cite{JML}, \cite{MAb}, and give an independent determination  of the ratio  $f_+^{DK}(0)/f_+^{D\pi}(0) $, which was considered a new independent parameter.

Finally from semileptonic decays $\bar{B}\rightarrow D l\nu$  and  $\bar{B}\rightarrow D^* l\nu$, \cite{BA5}, \cite{BA6}, \cite{BA7}, \cite{BA}, \cite{IA},  \cite{BA9}, \cite{BA10}, \cite{JAb}, \cite{NEA1}, \cite{KAb}, \cite{PA},  \cite{GA}, one find  $V_{cb}, \;\mathcal{G}(1)$ and  $\mathcal{F}(1)$ parameters.

\section{Numerical results}

Data from  the above cited papers were used to define  $\chi_2^2$, the second component of full $\chi^2$, which has a parabolic form in $V_{ij}^2$.  The first component $\chi_1^2$,  (\ref{chi1}), that contain all unitarity constraints, has a  parabolic part, and one that is highly nonlinear in all $V_{ij}$. Thus we had to test the stability of the expected physical values against the strong non-linearity implied by unitarity.
Eventually the chosen method was to modify all the measured central values in the same sense, plus and minus, respectively, proportional to  their corresponding uncertainties. 

An important assumption included in  our approach is that   numerical $V_{ij}$ values must be the same irrespective of the physical processes used to determine them.  Accordingly  the other parameters, such as decay constants, $f_P$, form factors,  $f_+(0)$, $\lambda$, etc., that parametrize each given experiment, have been considered as independent  parameters to be obtained from  fit, by  applying the usual technique to obtain their mean values and uncertainties. 

The stability tests provided sets of different moduli matrices that have been
 used to  obtain a mean value matrix and its corresponding error matrix. The mean and uncertainty matrices have been  computed by embedding  unitary matrices into the double stochastic matrix set, see \cite{PD}.

The central values and uncertainties of data used in fit are those published in the above cited papers, and we combined the statistical and systematic  uncertainties in quadrature when experimentalists provided both of them.

The numerical values obtained  from fit for  decay constants, $f_{\pi},\,f_K,\,f_B,\, f_D,\, f_{D_s}$, semileptonic form factors $f_+(0)$,  $\Delta_R^V,\; \lambda$, as well as the ratios $f_K/f_{\pi}\; {\rm and},\; f_+^{DK}(0)/f_+^{D\pi}(0)$ are given in TABLE I. All of them are in the expected range although many have big uncertainties. 

\begin{table}
\caption{\label{tab:table1} Numerical values for decay constants $f_P$ and form  factors $f_+(0)$ in MeV units, and $\Delta_R^V$ and $\lambda$}
\begin{ruledtabular}
\begin{tabular}{llllll}
Parameter&Central Value& Uncertainty&Parameter&Central Value& Uncertainty\\
\hline
$f_{\pi}$ & 131.131&1.522&$f_+^{D \pi}(0)$&653.2&19.1\\
$f_K$& 154.97 & 2.17&$f_+^{D K}(0)$&751.8&10.4\\
$f_K/f_{\pi}$ & 1.1818 & 0.0042&$f_+^{D K}(0)/f_+^{D \pi}(0)$\footnotemark[1]&1.171&0.049\\
$f_B$&222.8&25.0&$\mathcal{F}$(1)&957.5&57.7\\
$f_D$&207.6&9.8&$\mathcal{G}$(1)&1,125.3&40.7\\
$f_{D_s}$&271.0&18.0&$\Delta_R^V\%$&2.373&0.096\\
$f_+^{K\pi}(0)$&955.34&9.27&$\lambda$&-1.2686&0.0057\\
$f_+^{B \pi }(0)$&214.9&13.4&$C_r$&$0.002$&$0.0001$\\
\end{tabular}
\end{ruledtabular}
\vskip2mm
\footnotemark[1] {Data on $V_{cd} f_+^{D\pi}/V_{cs} f_+^{D K} $ from  papers  \cite{SD}-\cite{MAb}},
\end{table}

Our approach allows a  ``fine structure analysis'' of all experiments measuring one definite quantity, such as $\Delta_R^V$, or $f_+^{\pi K}(0)$ . 

For example  KLOE collaboration data, \cite{Kl}, and FlaviaNet Working Group data, \cite{Fl}, on $f_+(0)V_{us}$ lead  to
\begin{eqnarray}
f_+^{\pi K}(0)_{KLOE}&=& 950.38 \pm 5.56,\; {\rm and} \nonumber\\
 f_+^{\pi K}(0)_{Flavia}&=& 955.06 \pm 4.31,\end{eqnarray}
respectively, whose central values are a little bit different, but compatible between them at $1\sigma$, and all together provide
\begin{eqnarray}
f_+^{\pi K}(0)_{KLOE+Flavia}=952.72 \pm 5.30\end{eqnarray}
$f_+^{\pi K}(0)$ from Table I has a precision of 1\%, and the above value obtained from  ten measurements has a precision of 0.56\%.

In this approach one could obtain  information about lepton universality in leptonic decays. Because the corresponding $f_M$ decay constant should be same for both decays $M\rightarrow \mu \nu$ and  $M\rightarrow \tau \nu$, a big difference between them could show a possible violation. As an example we chose $M = D_s^+$ meson since  lattice computations provided a number for this decay constant,   $f_{D_s}=241(3)$, see \cite{EF}, with a   very small error. Our results are 
\begin{eqnarray}
f_{D_s^+\rightarrow  \mu \nu} = 265.0\pm 14.0 \nonumber\\
f_{D_s^+\rightarrow  \tau \nu} = 276.0\pm 20.0\end{eqnarray}
which are consistent with lepton universality. The above two numbers together with that from TABLE I completely disagree with that provided by lattice computations, being far away from theoretical prediction at $8\sigma$,  $13\sigma$,  and $10\sigma$, respectively, 
 where, $\sigma=3$, is lattice uncertainty. The experimental spreading is, $246.0 \le  f_{D_s^+}\le 311.0$, and the minimal and maximal values correspond to the branching ratios obtained for  $\mathcal{B}(D_s^+\rightarrow \mu^+ \nu_{\mu})=(5.15 \pm 0.63 \pm 0.20 \pm 1.29)\times 10^{-3}$, see \cite{BA4}, and to the branching ratio  $\mathcal{B}(D_s^+\rightarrow \tau\nu)=(8.0 \pm 1.3 \pm 0.6)\%$,  given by Eq.(6) in paper \cite{AR1}, respectively. On the other hand if one computes the difference between each one of the above three values and that provided by lattice computation, divided by the corresponding experimental error $\sigma$  obtained from  fit one finds the same value, $1.17$, which shows again the consistency of  experimental data. In our opinion the lattice number is much  underestimated. In contradistinction their value for $f_K/f_{\pi} = 1.189(7)$ is not far from the fit value $f_K/f_{\pi} = 1.1818(42)$.
\begin{table}
\caption{\label{tab:table2} $\Delta_R^V$ central values and uncertainties provided by fit when using data from Refs. \cite{HT1}-\cite{GS}}
\begin{ruledtabular}
\begin{tabular}{cccccc}
Ref.&\cite{HT1}&\cite{HT2}&\cite{HT3}&\cite{GS}&\cite{HT2}\\
$\Delta_R^V\%$&2.373(96)&2.399(112)&3.361(196)&2.294(136)&2.361(38)\footnotemark[1]\\
\end{tabular}
\end{ruledtabular}
\vskip2mm\footnotemark[1]
{The ``constant'' $\Delta_R^V$ and its uncertainty, from \cite{HT2}}
\end{table}

Another unexpected  result concerns $\Delta_R^V$ constancy, usually assumed in all  four papers \cite{HT1}-\cite{GS}, assumption that is not  confirmed by our analysis. However our result $\Delta_R^V=2.373 \pm 0.096$ obtained from data \cite{HT1} is in   good concordance with  the value given by relation  (\ref{rad}), see TABLE II, but   our uncertainty    is $2.5 \sigma$ higher  than the theoretical one. In this case  $\Delta_R^V$  spreading that results from paper \cite{HT1} is,  
$2.2027 \le\Delta_R^V\le 2.472$, which corresponds to $7.1 \sigma$  where $\sigma=0.038\%$ is the theoretical uncertainty. The extremal nuclei are   $^{22}$Mg and $^{54}$Co, respectively. The spreading obtained from Savard \textit{et al} data, \cite{GS}, corresponds to  $10.8 \sigma$, and the extremal nuclei are $^{74}$Rb and $^{34}$Cl. However the difference between the central results from \cite{HT1} and \cite{GS} is 2$\sigma$, which suggests that there is still room for computation improvements.  The  mean values and the corresponding uncertainties for all data from papers \cite{HT1}-\cite{GS} are given in TABLE II.

\begin{table}
\caption{\label{tab:table3} $|f_+(q^2)|$ form  factors  from 
 $D\rightarrow \pi l \nu$  and $D\rightarrow K l \nu $ decays}
\begin{ruledtabular}
\begin{tabular}{ccccc}
Bin& $f_+^{\pi}(q^2)\;(\pi^{0}e^+\nu_e)$&$ f_+^{\pi}(q^2)\;(\pi^-e^+\nu_e)$& $f_+^{K}(q^2)\;(K^{-}e^+\nu_e)$&$ f_+^{K}(q^2)\;(\bar{K}^0e^+\nu_e)$\\
\hline
1&0.6895 $\pm$ 0.0576&0.7072 $\pm$ 0.0355&0.7798 $\pm$ 0.0136&0.7808 $\pm$ 0.0208 \\
2&0.7514  $\pm$ 0.0709&0.7735 $\pm$ 0.0400&0.8281 $ \pm$ 0.0146& 0.8096 $\pm$ 0.0239\\
3&0.8442 $\pm$ 0.0841&0.7956 $\pm$ 0.0488&0.8435 $ \pm$ 0.0156& 0.8466  $\pm$ 0.0259\\
4&0.8926  $\pm$ 0.0974&0.9812 $\pm$ 0.0532&0.9113  $ \pm$ 0.0198&0.8856  $\pm$ 0.0289\\
5&1.1005  $\pm$ 0.115&1.017  $\pm$ 0.065&0.9945  $ \pm$ 0.0229&0.9175 $\pm$ 0.0331\\
6&1.3083 $\pm$ 0.151&1.101  $\pm$ 0.084&1.007  $ \pm$ 0.027&1.024 $\pm$ 0.039\\
7&1.5789  $\pm$ 0.221&1.635  $\pm$ 0.111&1.128  $\pm$ 0.033&1.149  $\pm$ 0.048\\
8&-&1.759  $\pm$ 0.235&1.212 $\pm$ 0.042 &1.090  $\pm$ 0.062\\
9&-&2.024  $\pm$ 0.301&1.303 $\pm$ 0.066&1.233  $\pm$ 0.092\\
10&-&-&1.561 $\pm$ 0.165&1.476 $\pm$ 0.210\\
\end{tabular}
\end{ruledtabular}
\end{table}

Our approach allows the form factors determination. The paper \cite{JG} provided for the first time results on $V_{cq} |f_+(q^2)|$, $q=d,\,s$, and in TABLE III 
are given our determinations. Their graphic form  is shown  in FIG.\,1 at the end of paper.

The precision of $f_P$ and $f_+(0)$ determinations varies from 1\% for $f_+^{K\pi }(0)$ to 11\% for $f_B$. More about variability of the above parameters could be  learnt  from KM moduli matrix. Our fit result for  KM  central moduli values is 

\begin{eqnarray}
V_c=\left(\begin{array}{lll}
0.974022&0.226415&0.0042512\\
0.226253&0.973323&0.0381075\\
0.0095692&0.0371307&0.999265\end{array}\right)\label{centr}
\end{eqnarray}
where the digit numbers are those suggested by Mathematica rounding, when working in double precision. Its associated uncertainty matrix is
\begin{eqnarray}
\sigma{_{V_c}}=\left(\begin{array}{lll}
1.1\times 10^{-6}&1.9\times 10^{-5} &2.1\times 10^{-5} \\
3.6\times 10^{-5} &2.7\times 10^{-4} &3.1 \times 10^{-4}\\
3.5\times 10^{-5} &2.9\times 10^{-4} &3.9\times 10^{-4} \end{array}\right)\label{sig}
\end{eqnarray}

The last matrix has been obtained with the help of stabilty tests. One such  matrix is (\ref{sig1}) that was obtained when all the central measured values have been modified  with plus one tenth from the corresponding uncertainty. Although such a modification is highly improbable from an experimental point of view, it brings to light the variation direction for all parameters entering the fit.
\begin{eqnarray}
V_+=\left(\begin{array}{lll}
0.974021&0.226332&0.0074844\\
0.226114&0.973083&0.044512\\
0.0124317&0.043391&0.998891\\
 \end{array}\right)\label{sig1}
\end{eqnarray}

For example the above three matrices show that $V_{ud}$ is precisely determined with five digits, while  $V_{us},\;V_{cd},$ and $V_{cs}$ only with  three digits.
 $V_+$ matrix also shows the high $V_{ub}$ volatility, such that future data could lead to higher values for it than that given by $V_c$ matrix. In fact matrices $V_+$ and $V_c$ show that $f_B \in (123.5, 222.8)$. Although our value for
$V_{ub}=(4.25 \pm 0.02)\times 10^{-3}$ is higher than that from  PDG fit, \cite{pdg08}, it is compatible with that obtained in  Ref. \cite{BA3}, $V_{ub}= (4.1 \pm 0.2_{st} \pm 0.2_{syst}\,^{+0.6}_{-04.FF})\times 10^{-3}$.

 The new data from $D$ leptonic, \cite{BE}-\cite{ID}, and semileptonic decays, \cite{DCH}-\cite{JG},  combined with the new data from $\bar{B}\rightarrow Dl\nu$  and  $\bar{B}\rightarrow D^*l\nu$ \cite{BA5}-\cite{IA}, changed the KM moduli values, in particular those from the last row and column, see  \cite{pdg08}, p. 150. 

A big step forward will be the  measurement of  $q^2$ dependence for products of the  form $|f_+(q^2)V_{qb}|$, where $q=u,\,c$, for $B\rightarrow \pi l \nu$, $\bar{B}\rightarrow Dl\nu$  and  $\bar{B}\rightarrow D^*l\nu$ decays, similar   to that done for $D$ semileptonic decays. The simplest case is that of  $|f_+(q^2)V_{ub}|$ because there are measured data in twelve bins wich can be transformed in values for the product $|f_+(q^2)V_{ub}|$, \cite{BA3}. Such measurements will allow a more precise  $V_{ub}$ and  $V_{cb}$ determination, and, by consequence, it will provide better values for all KM matrix moduli from the the first two rows, 
and, perhaps, the  first hints for  new physics beyond SM, if any.
 
 $V_c$ matrix provides numerical values for $\delta$ and  angles of the standard unitarity triangle, as follows
\begin{eqnarray}\begin{array}{cccc}
\delta &=& (89.96\, \pm\, 0.36)^{\circ},&\alpha=(64.59 \, \pm\, 0.27)^{\circ},\\
\gamma&=&(89.98 \,\pm\, 0.06)^{\circ},&\beta=(25.49 \, \pm\,0.28)^{\circ}\end{array}\end{eqnarray}

The $\delta$ value could be interpreted as a maximal violation of  {\em CP} symmetry because $\sin\delta \approx 1$. A surprising change is the shape of 
 the standard unitarity triangle that is now  a right triangle, since  $\gamma\approx 90^{\circ}$. 
The Jarlskog invariant is 
\begin{eqnarray}J=(3.567 \pm 0.007)\times10^{-5}\end{eqnarray}
Similar results are obtained from  $V_+$ matrix even if $V_{ub}$ is almost twice bigger than that from $V_c$; they are
\begin{eqnarray}\begin{array}{cccc}
\delta &=& (90.0\, \pm\, 0.2)^{\circ},&\alpha=(54.1 \, \pm\, 0.1)^{\circ},\\
\gamma&=&(90.0 \,\pm\, 0.4)^{\circ},&\beta=(36.0 \, \pm\,0.2)^{\circ}\end{array}\end{eqnarray}

The above results show that $\delta$ and $\gamma$ angles are independent of $V_{ub}$ variation, while $\alpha$ and $\beta$ are moderately dependent. Thus experimental results on  the product $|f_+(q^2)V_{ub}|$ could lead to the determination of all the angles of the standard unitarity triangle.

The small angles uncertainties show that $V_c$ and $V_+$ matrices are higly compatible with unitarity constraints and all 165 $\cos\delta$ formulas provide very close each other values for all of them.

A comparison with lattice computations from \cite{CA} shows that $f_+^{D\pi}(0)$ and  $f_+^{DK}(0)$, as well as their ratio are in good agreement with the exprimental values obtained by us, and our uncertainties are smaller. An open problem remains  lattice computations of $f_B$ and  $f_+^{B\pi}(0)$ parameters.

\vskip.5cm
\begin{center}
\begin{figure}[h]
\begin{tabular}{cc}
{\includegraphics[height=2.2in]{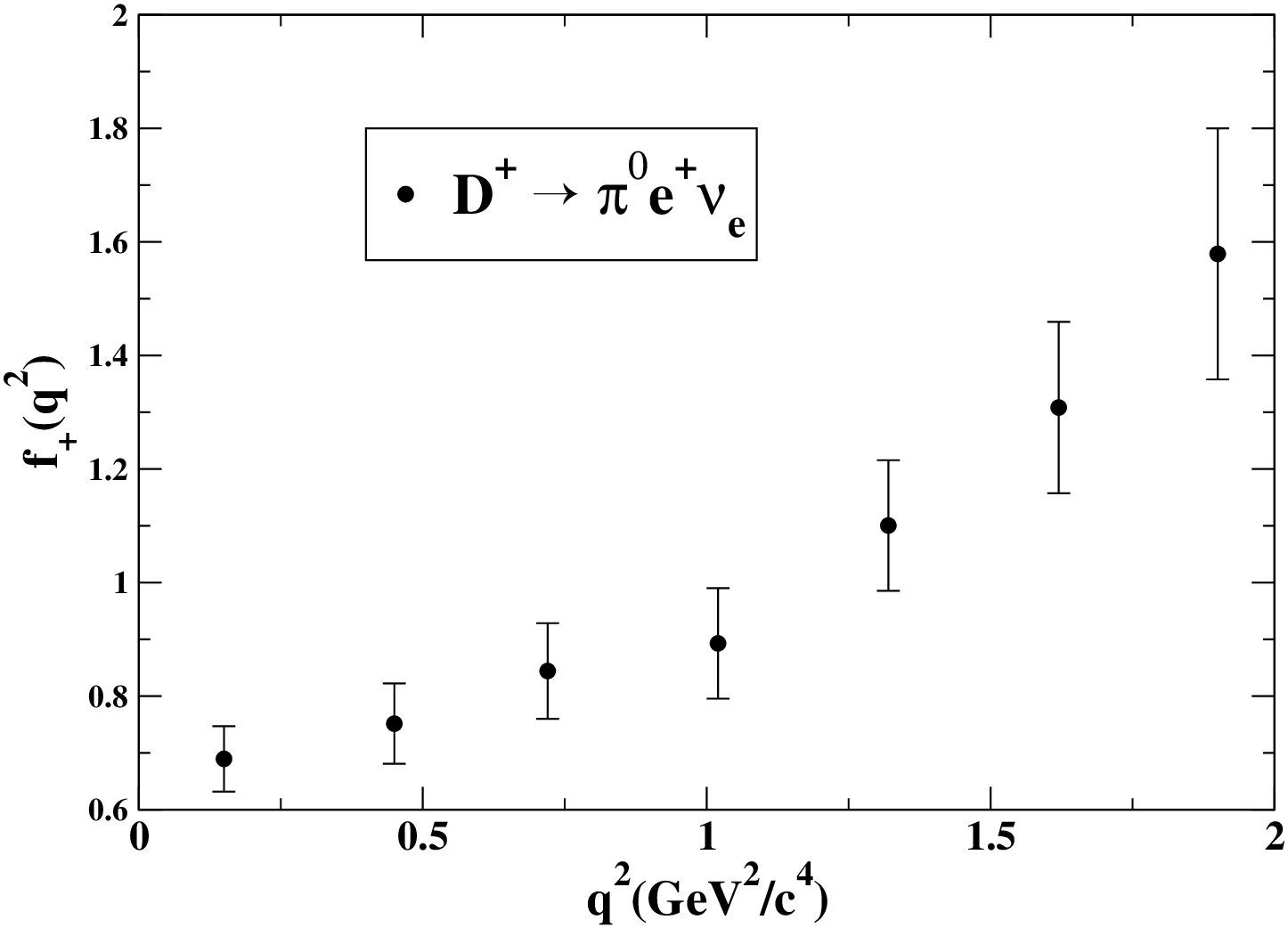}}~~~~&~~~~
{\includegraphics[height=2.2in]{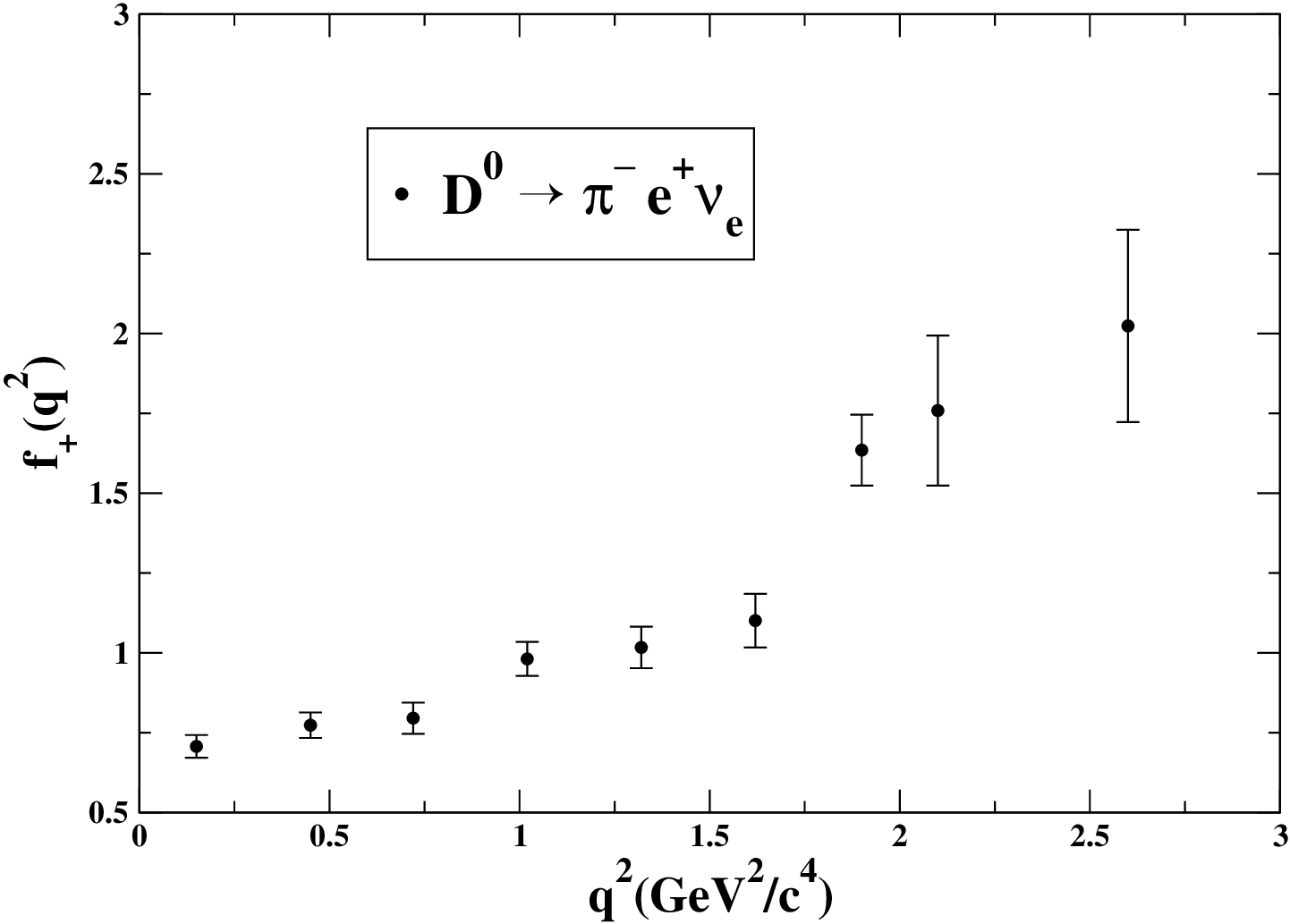}}\end{tabular}
\end{figure}
\vskip0.5cm

\begin{figure}[h]
\begin{tabular}{cc}
{\includegraphics[height=2.2in]{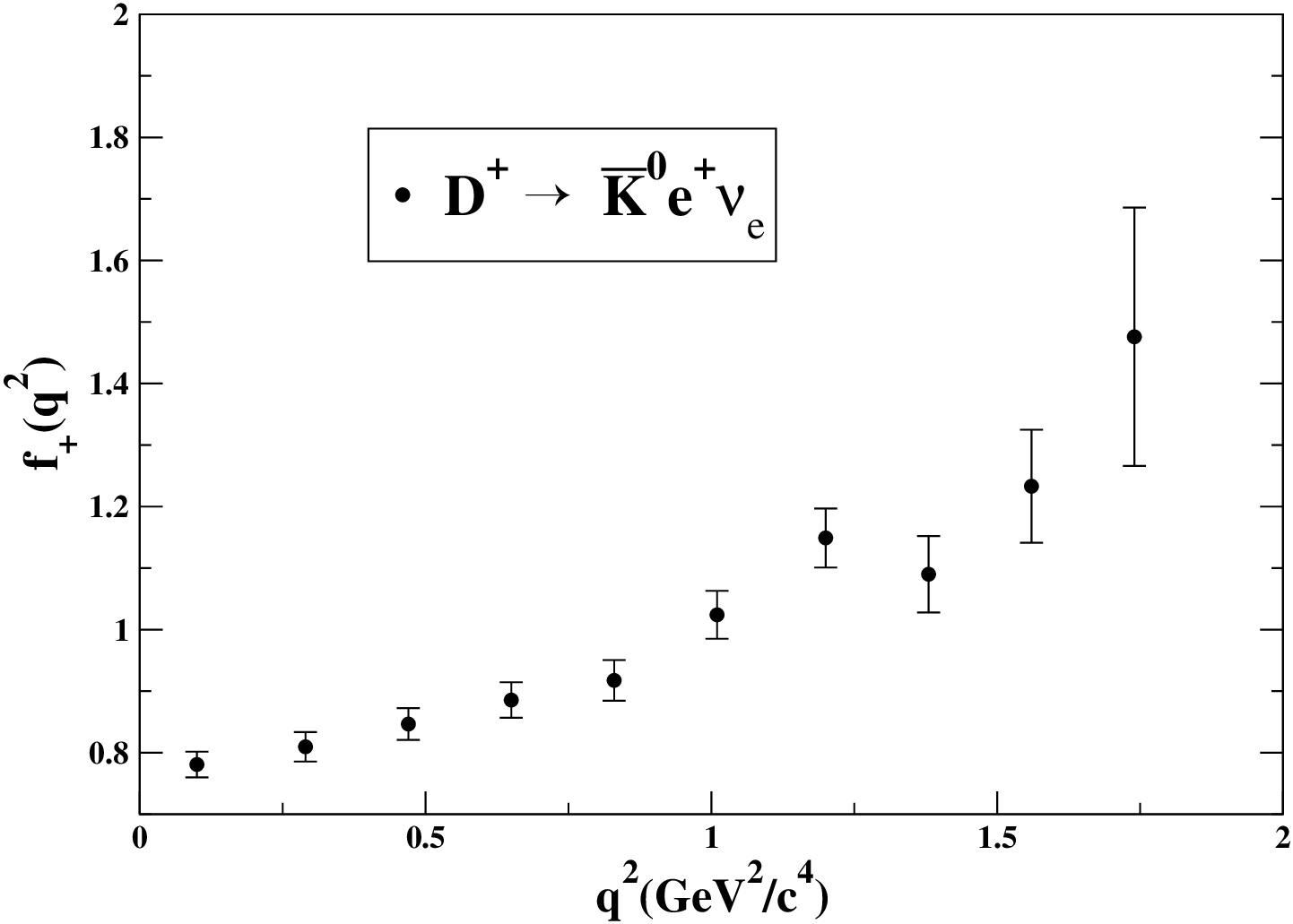}}~~~~&~~~~
{\includegraphics[height=2.2in]{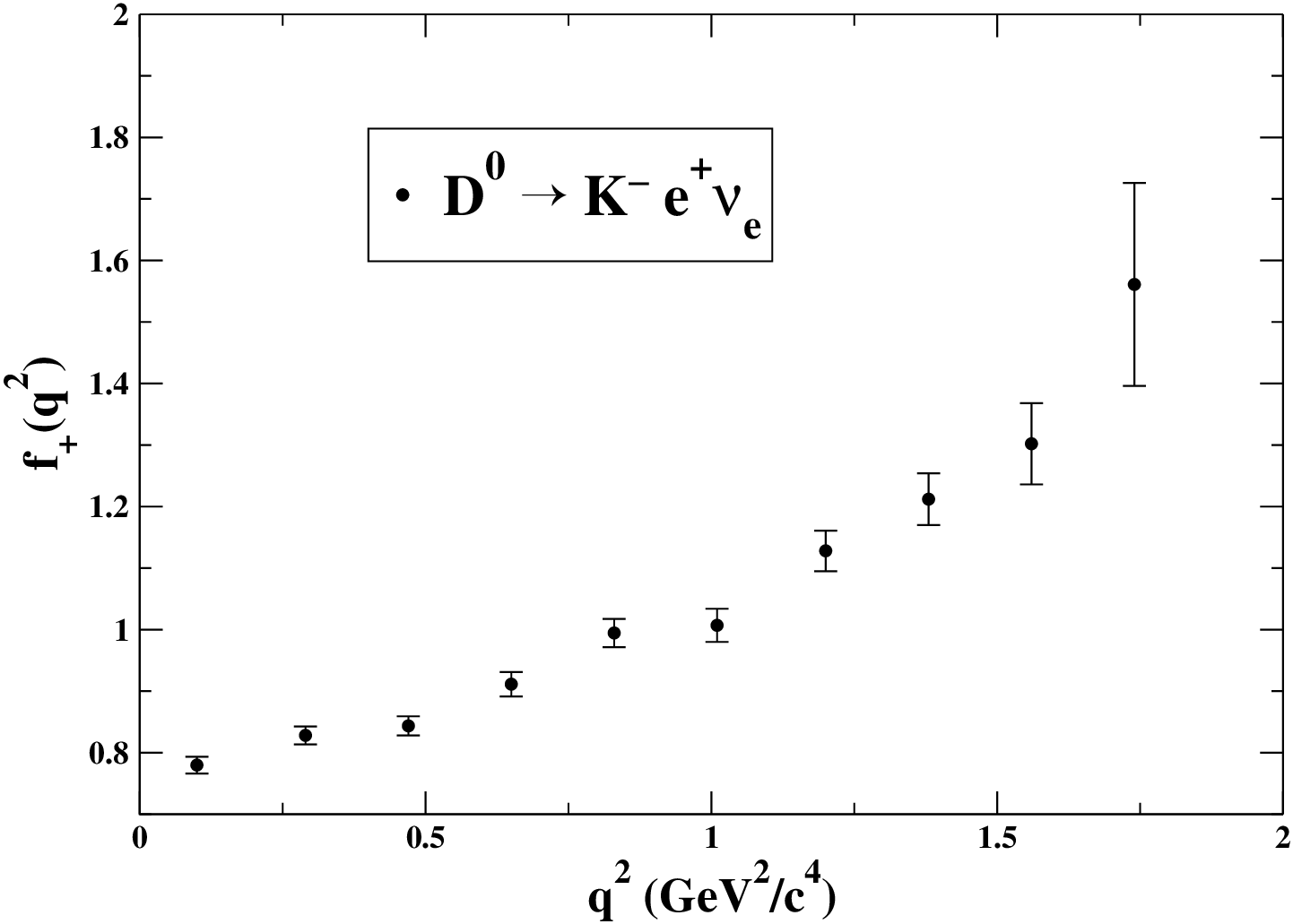}}\end{tabular}
\end{figure}
FIG. 1\; $|f_+(q^2)|$ Form  Factors from $D\rightarrow \pi l \nu $
and $D\rightarrow K l \nu $  decays 
\end{center}


 \section{Conclusion}

In this paper we presented a phenomenological tool that allows determination of KM moduli, semileptonic form factors and  decay constants  directly from experimental data. It is based on a  rephaising invariant  implementation of unitarity constraints that makes use of KM matrix moduli as fit parameters. These  constraints  are strong enough and give   a consistent picture of nowadays flavor physics, and until now provide  no signals for new physics beyond the SM.
However there is a discrepency between theoretical $f_{D_s^+}$ lattice computation value  for the  $D_s^+\rightarrow l\nu$ decay constant, and the experimental value obtained from  fit, that suggests that the numerical value is highly underestimated.

A feature of our tool is that all measurable parameters are each other strongly correlated, a little modification of one of them propagates to all the other parameters, property which is a consequence of unitarity constraints.

The new data from $D$ leptonic and semileptonic decays, and those from  $\bar{B}\rightarrow D(D^*)l\nu$ led to a significant change of KM moduli and of standard  unitarity triangle shape.
 A crucial step forward would be the measurement of $B\rightarrow \pi l \nu$ form factors  in bins of $0.5$ GeV$^2$ that will allow a better $V_{ub}$ determination, and a stabilization of moduli values entering the first two rows.

Taking into account that our approach gives reliable results for many   parameters  entering  flavor physics an important task comes to the lattice community to improve their numerical algorithms, because detection of  new physics could be 
the outcome of both experimentalists and theorists.

\vskip.5cm

{\bf Acknowledgements.}  We acknowledge a partial support from ANCS Contract No 15EU/06.04.2009.

\vskip0.5cm
 Electronic address: dita@zeus.theory.nipne.ro

\begin{thebibliography}{99}

\bibitem{KM} M. Kobayashi and T. Maskawa, Progr. Theor. Phys. \textbf{ 49}, 652 (1973)
\bibitem{pdg08} C.Amsler  \textit{at al.}, (Particle Data Group), Phys. Lett. B \textbf{ 667}, 1 (2008)
\bibitem{HL} A. H\"ocker and Z. Ligeti, Annu. Rev. Nucl. Part. Sci. \textbf{ 56}, 501 (2006); 
see p. 507 bottom.
\bibitem{J1} C. Jarlskog, Z.Phys. C \textbf{ 29}, 491 (1985); Phys. Rev. D \textbf{ 35}, 1685 (1987)
\bibitem{J2} C. Jarlskog, Phys. Rev. Lett. \textbf{ 55}, 1039 (1985)

\bibitem{AKL} R.Aleksan, B. Kayser, and D. London, Phys. Rev. Lett. \textbf{ 73}, 18 (1994)
\bibitem{BL} G.C. Branco and L. Lavoura, Phys. Lett. \textbf{ 208}, 123 (1998)
\bibitem{J3} C. Jarlskog and R. Stora, Phys. Lett. \textbf{ 208}, 268 (1998)
\bibitem{PD} P. Di\c t\u a, J. Math. Phys. \textbf{ 47},  083510 (2006)
\bibitem{HT1} J. C. Hardy and I. S. Towner,  Phys. Rev. C \textbf{79}, 055502 (2009)
\bibitem{HT2} J. C. Hardy and I. S. Towner, Phys. Rev. C \textbf{77}, 025501 (2008)
\bibitem{HT3} J. C. Hardy and I. S. Towner, Phys. Rev. C \textbf{71}, 055501 (2005)
\bibitem{GS} G Savard  \textit{et al}, Phys. Rev. Lett. \textbf{ 95} 102501 (2005)
\bibitem{I} T. M. Ito, arXiv:0704.2365

\bibitem{CLN} I. Caprini, L. Lellouch, and M. Neubert, Nucl. Phys. \textbf{ B350}, 153 (1998)
\bibitem{AS} A. Serebrov \textit{et al}, Phys. Lett. B  \textbf{ 605} 72 (2005)
\bibitem{JN} J. S. Nico \textit{ et al}, Phys. Rev.  C \textbf{ 71} 055502 (2005)
\bibitem{MD} M. S. Dewey \textit{et al}, Phys. Rev. Lett. \textbf{ 91} 152302 (2003)
\bibitem{SAr} S. Arzumov \textit {et al}, Phys. Lett. B  \textbf{ 483} 15 (2000)
\bibitem{RP} R. W. Pattie, Jr.  \textit { et al}, Phys. Rev. Lett. \textbf{ 102} 012301 (2009)
\bibitem{HA} H. Abele, Progr. Part. Nucl. Phys. \textbf{ 60}, 1 (2008)
\bibitem{HA1} H. Abele  \textit{ et al}, Phys. Rev. Lett. \textbf{ 88} 211801 (2002)
\bibitem{BY} B. Yerozolimsky  \textit{ et al}, Phys. Lett. B  \textbf{ 412} 240 (1997)
\bibitem{JB}J. Byrne   \textit{ et al}, J. Phys.G.: Nucl. Part. Phys. \textbf{ 28}, 1325 (2002)
\bibitem{WJM} W. J. Marciano, Phys. Rev. Lett. \textbf{ 93} 231803 (2004)
\bibitem{Kl} F.Ambrosino    \textit{ et al}, (KLOE Collaboration), JHEP04,\textbf{ 059} (2008)
\bibitem{Fl} M. Antonelli   \textit{ et al}, (FlaviaNet Kaon Working Group),  arXiv:0801.1817
\bibitem{MT} M. Testa, arXiv:0805.1969v1

\bibitem{JR} J. R. Batley  \textit { et al}, (NA48/2 Collaboration), Eur. Phys. J. C \textbf{ 50}, 329 (2007)
\bibitem{CB} C. Bloise, Acta Phys.Pol. B \textbf{ 38} 2731 (2007)
\bibitem{FA} F. Ambrosino  \textit{ et al}, arXiv:0701008v2

\bibitem{Ts} T. Spadaro, arXiv:hep-ex/0703033v1
\bibitem{Vr} V. I. Romanovski  \textit{ et al}, arXiv:0704.2052v1 [hep-ex]
\bibitem{Mm} M. Moulson, arXiv: hep-ex/0703013v1 
\bibitem{FA2} F.Ambrosino  \textit { et al},  (KLOE Collaboration), Phys. Lett. B  \textbf{ 632} 43 (2006)

\bibitem{FA3} F.Ambrosino  \textit { et al}, (KLOE Collaboration),
Phys. Lett. B  \textbf{ 636} 173 (2006)

\bibitem{MT1} M. Testa, Nucl.Phys. B(Proc. Suppl) {\bf 162} 205 (2006)

\bibitem{AB}  A. Bevan \textit{ et al}, (NA48/2 Collaboration), Phys. Lett. B  {\bf 602} 41 (2004)
\bibitem{TA} T. Alexopoulos  { et al.} (KTeV Collaboration), Phys. Rev. Lett. {\bf 93} (2004) 181802

\bibitem{SA} A. Sher {\em et al.}, Phys. Rev. Lett. {\bf 91},  261802 (2003)
\bibitem{SBA} S. B. Athar  \textit{ et al},(CLEO Collaboration),  Phys. Rev. D {\bf 68}, 072003 (2003)
\bibitem{TH} T. Hokuue \textit{ et al}, (Belle Collaboration), Phys. Lett. B  {\bf 648} 139 (2007)

\bibitem{KI} K. Ikado  \textit{ et al.}, (Belle Collaboration), Phys. Rev. Lett. {\bf 97},  251802 (2006)
\bibitem{NEA} N. E. Adam \textit{ et al}, (CLEO Collaboration), Phys. Rev. Lett. {\bf 99},  041802 (2007)
\bibitem{BA1} B. Aubert  \textit{ et al}, (BaBar Collaboration), Phys. Rev. D {\bf 76}, 052002 (2007)
\bibitem{BA2} B. Aubert \textit { et al}, (BaBar Collaboration), Phys. Rev. D {\bf 77}, 011107 (2008)

\bibitem{FN} J.M. Flynn and J. Nieves,  Phys. Rev. D {\bf 76}, 031302 (2007)
\bibitem{BA3} B. Aubert  \textit{ et al}, (BaBar Collaboration), Phys. Rev. Lett. {\bf 98}, 091801 (2007)
\bibitem{BLC} C. Bourrely, L. Lelouch, and I. Caprini,  Phys. Rev. D {\textbf 79}, 013008 (2009)
\bibitem{BE} B. I. Eisenstein   \textit{ et al}, (CLEO Collaboration), Phys. Rev. D {\bf 78}, 052003 (2008)
\bibitem{AR} M. Artuso \textit  \textit{ et al}, (CLEO Collaboration), Phys. Rev. Lett. {\bf 95},  251801 (2005)

\bibitem{ID} I. Dank\'o, J.Phys.Conf.Series {\bf 9}, 91 (2005)
\bibitem{JPA} J. P. Alexander   \textit{ et al},  (CLEO Collaboration), Phys. Rev. D {\bf 79}, 052001 (2009)
\bibitem{PU} P. U. E. Onyisi  \textit{ et al}, (CLEO Collaboration), Phys. Rev. D {\bf 79}, 052002 (2009)
\bibitem{KME} K.M. Ecklund  \textit{ et al}, (CLEO Collaboration),  Phys. Rev. Lett. {\bf 100}, 161801 (2008)
\bibitem{LW} L. Widhalm  \textit{ et al}, (Belle Collaboration),  Phys. Rev. Lett. {\bf 100}, 241801 (2008)
\bibitem{AR1} M. Artuso   \textit{et al}, (CLEO Collaboration), Phys. Rev. Lett. {\bf 99}, 071802 (2007)
\bibitem{BA4} B. Aubert  {\em et al}, (BaBar Collaboration), Phys. Rev. Lett. {\bf 98}, 141801 (2007)
\bibitem{DCH} D. Cronin-Henessy  \textit{ et al}, (CLEO Collaboration),  Phys. Rev. Lett. {\bf 100}, 251802 (2008)
\bibitem{SD} S. Dobbs  \textit{ et al}, (CLEO Collaboration),  Phys. Rev. D {\textbf 77}, 112005 (2008)
\bibitem{JG} J. Y. Ge \textit{ et al}, (CLEO Collaboration),  Phys. Rev. D {\textbf 79}, 0502010 (2008)
\bibitem{LW1} L. Widhalm \textit{ et al}, Phys. Rev. Lett. \textbf{ 97}, 061804 (2006)
\bibitem{GSH} G. S. Huang \textit{ et al}, (CLEO COllaboration),  Phys. Rev. Lett. {\bf 94}, 011802 (2006)
\bibitem{JML} J.M. Link \textit{ et al} (FOCUS Collaboration), Phys. Lett. {text\bf B 607}, 51 (2005)
\bibitem{MAb} M. Ablikim \textit{ et al}, (BES Collaboration), Phys. Lett. B  \textbf{ 597} 39 (2004)
\bibitem{BA5} B. Aubert \textit { et al}, (BaBar Collaboration), Phys. Rev. D \textbf{ 79}, 012002 (2009)
\bibitem{BA6} B. Aubert  \textit{ et al}, (BaBar Collaboration), arXiv:0904.4063v1
\bibitem{BA7} B. Aubert \textit { et al}, (BaBar Collaboration), Phys. Rev. Lett. \textbf{ 100}, 231803 (2008)
\bibitem{BA} B. Aubert  \textit{ et al},  (BaBar Collaboration), Phys. Rev. D \textbf{ 77}, 032002 (2008)
\bibitem{IA} I. Adachi  \textit{ et al},(BELLE Collaboration),  Phys. Rev. D \textbf{ 78}, 032016 (2008)
\bibitem{BA9} B. Aubert  \textit{ et al}, (BaBar Collaboration), arXiv:0712.3493v2
\bibitem{BA10} B. Aubert \textit { et al}, (BaBar Collaboration), Phys. Rev. D \textbf{ 71}, 051502 (2005)
\bibitem{JAb} J. Abdallah  \textit{ et al.} (DELPHI Collaboration), Eur. Phys. J C \textbf{ 33} 21 (2004)
\bibitem{NEA1} N. E. Adam \textit{ et al}, (CLEO Collaboration), Phys. Rev. D \textbf{ 67}, 032001 (2003)
\bibitem{KAb} K. Abe \textit{ et al}, (BELLE Collaboration), Phys. Lett. B 
 \textbf{ 526} 247 (2002)
\bibitem{PA} P. Abreu \textit{ et al}, (DELPHI Collaboration), Phys. Lett. B 
 \textbf{ 510} 55 (2001)
\bibitem{GA} G. Abbiendi  \textit{ et al}, (OPAL Collaboration), Phys. Lett. B  \textbf{ 482} 15 (2000)
\bibitem{EF} E. Follana, C. T. H. Davies, G. P. Lepage, and J. Shigemitsu,  Phys. Rev. Lett., \textbf{ 100}, 062002 (2008)
\bibitem{CA} C. Aubin  \textit{ et al}, (Fermilab Lattice Collaboration, MILC  Collaboration, and HPQCD  Collaboration), Phys. Rev. Lett. \textbf{ 94}, 011601 (2005)
\end{thebibliography}
\end{document}